\documentclass[12pt,preprint]{aastex}
\begin{document}

\title{Beryllium in Ultra-Lithium-Deficient Halo Stars -- \\ 
The Blue Straggler Connection}

\author{Ann Merchant Boesgaard\altaffilmark{1}}
\affil{Institute for Astronomy, University of Hawai`i at M\-anoa, \\
2680 Woodlawn Drive, Honolulu, HI {\ \ }96822 \\ }
\email{boes@ifa.hawaii.edu}

\altaffiltext{1}{Visiting Astronomer, W.~M.~Keck Observatory jointly operated
by the California Institute of Technology and the University of California.}

\begin{abstract}
There exists a small group of metal-deficient stars that have Li abundances
well below the Li plateau that is defined by over 100 unevolved stars with
temperatures above 5800 K and values of [Fe/H] $<$ $-$1.0.  Abundances of Be
have been determined for most of these ultra-Li-deficient stars in order to
investigate the cause of the Li deficiencies.  These Li-deficient stars have
implications on the value of primordial Li.  High-resolution and high
signal-to-noise spectra have been obtained in the Be II spectral region near
3130 \AA\ for six ultra-Li-deficient stars with the Keck I telescope and its
new uv-sensitive CCD on the upgraded HIRES.  The spectrum synthesis technique
has been used to determine Be abundances.  All six stars are found to have Be
deficiencies also.  Two have measurable - but reduced - Be and four have only
upper limits on Be.  These results are consistent with the idea that these Li-
and Be-deficient stars are analogous to blue stragglers.  The stars have
undergone mass transfer events (or mergers) which destroy or dilute both Li
and Be.  The findings cannot be matched by the models that predict that the
deficiencies are due to extra-mixing in a subset of halo stars that were
initially rapid rotators, with the possible exception of one star, G 139-8.
Because the ultra-Li-deficient stars are also Be-deficient, they appear to be
genuine outliers in population of halo stars used to determine the Li plateau
in that they no longer have the Li in their atmospheres that was produced in
the Big Bang.
\end{abstract}


\section{Introduction}

The plateau in Li abundances was first noted by Spite \& Spite (1982).  Much
subsequent research work has added many more stars and has refined the details
of the plateau, e.g. Rebolo et al.~(1988), Hobbs \& Duncan (1987), Thorburn
(1994), Bonifacio \& Molaro (1997), Ryan et al.~(1999, 2001b), Novicki (2005)
and more recently Shi et al.~(2007) and Bonifacio et al.~(2007).  In the
course of these studies some Li ``misfits'' were discovered, the first of
which was HD 97916 found by Spite et al.~(1984); it had a low upper limit on
its Li content that was an order of magnitude below the plateau value.
Additional stars with only upper limits on the Li abundance were found by
Hobbs et al.~(1991), Hobbs \& Mathieu (1991), Thorburn (1994), and Ryan et
al.~(2001b).  They were dubbed ``ultra-lithium-deficient'' halo stars by Ryan
et al.~(2001a).  There are now nine such stars known with effective
temperatures higher than 5900 K.  (There are three others with upper limits
found by Thorburn (1994) at low [Fe/H], but the upper limits are not very
restrictive.)

The study of these Li-deficient halo stars and their properties is important
for two particular reasons: 1) what they can tell us about stellar structure
and evolution and 2) the assessment the amount of Li produced in the Big Bang.
If ``normal'' halo stars keep their original Li, what properties are different
in these stars that result in the loss of Li?  Are they merely examples of
extreme Li depletion and thus imply that other plateau stars are also
depleted, but more mildly?  If they need to be included in assessments of a
spread in Li due to a universal depletion, this would alter the value of the
Big Bang contribution to the overall Li abundance, log N(Li)$_p$ on the scale
where log N(H) = 12.00.

The latest predicted value of log N(Li)$_p$ that meets the requirement of
consistency with the results from the Wilkinson Microwave Anisotropy Probe
(WMAP) is 2.64 $\pm$0.03 according to Spergel et al.~(2006).  This is
considerably higher than the measured values in the most metal-poor stars;
recent results from Charbonnel \& Primas (2005), Novicki (2005), Boesgaard et
al.~(2005), Bonifacio et al.~(2007) give values of log N(Li)$_{obs}$ near 2.1.

Deliyannis et al.~(1990) considered the effects of diffusion and estimated
that it could decrease the initial Li abundance by 0.14 dex.  In their Figure
3 Ryan et al.~(1996) show the predicted effects of both rotation and diffusion
on the initial Li content compared to the extant Li abundances all assembled
on uniform temperature and abundance scales.  More recently, new observations
in the metal-poor globular cluster, NGC 6397, by Korn et al. (2006) suggest
that the observed Li reduction resulting from turbulent diffusion could be as
much as 0.25 dex to give A(Li) = 2.54 $\pm$0.10, thereby achieving agreement
with the WMAP results.  A recent intriguing idea has been put forward by Piau
et al.~(2006) that Big Bang Li was depleted by the first generations of stars.
They suggest that as much as one-half of the baryonic matter was effectively
depleted in Li and then recycled to the interstellar medium where efficient
turbulent mixing resulted in fairly constant level of Li in subsequent
generations of stars.  Then those stars may have further depleted Li by
internal mixing, e.g. diffusion, rotationally-induced mixing.

The connection of these very Li-deficient halo stars with blue stragglers has
been discussed in several papers, e.g.~Hobbs \& Mathieu (1991), Hobbs et
al.~(1991), Norris et al.~(1997), but the hypothesis of the ultra-Li-deficient
stars as blue-straggler analogs was put forward by Ryan et al.~(2001a, 2002)
for the whole subset of nine stars.  In addition, Carney et al.~(2001, 2005)
have done radial velocity studies of five of the metal-poor, Li-deficient
stars which are candidate field blue-stragglers.

There are at least two ideas about the cause of the Li deficiencies.  The
published ones involve blue-straggler analogs (Ryan et al.~2001a, 2002) and
rotationally-induced mixing (Pinsonneault et al.~1999, 2002).  These two
theories of the Li deficiency make different predictions about the Be content.
In the case of the blue-straggler analog, {\it all of most of the Be would be
destroyed}, while in the case of extra-mixing induced by stellar rotation {\it
all or most of the Be would be preserved.}

The blue stragglers in globular clusters stand out because they are blue and
they are above the main-sequence turn-off.  The blue-straggler analogs
under discussion here are thought to have extra mass deposited on them, but
are not brighter than their main-sequence turn-off counterparts.  The
phenomena of mass transfer or binary coalescence can affect lower mass,
sub-turn-off stars as well.  In a binary merger the thorough mixing would
destroy both Li and Be.  In mass transfer from an evolved star the Li and Be
would both have been diluted as the convection zone deepened; the mass
deposited on the ``analog'' would have little or no Li or Be present.

Pinsonneault et al.~(1999, 2000) suggest that a subset of rapid rotators
became the very-Li-deficient halo stars; as they spun down they would
lose Li by extra mixing.  Rotationally-induced mixing would transport the
surface Li atoms down to layers where the temperature is about 2.5 $\times$
10$^6$ K where they interact with a proton and become two helium nuclei.  The
mixing would have to be much more effective to destroy the surface Be atoms as
they would need to be mixed down to 3.5 $\times$ 10$^{6}$ K to be destroyed.

The most metal-poor of the ultra-Li-deficient stars is G 186-26.  Boesgaard \&
Novicki (2005) have determined its Be upper limit from UV spectra obtained at
Subaru with HDS.  They found that star to be ultra-Be-deficient also, with
A(Be) $<$$-$2.00.  In comparison with other stars at [Fe/H] $\sim$$-$2.7, its
{\it upper limit} on Be is 0.80 dex lower than normal Be.  The maximum amount
of Be depletion that would be caused by rotationally-induced mixing is 0.40
dex, according to the model calculations of Pinsonneault et al.~(1992).
Boesgaard \& Novicki (2005) suggested that the blue straggler analog could
explain the Li and Be upper-limit abundances via a binary merger or mass
transfer.

In this paper Be results for six more of the nine ultra-Li-deficient halo
stars are presented.  High resolution, high signal-to-noise spectra have been
observed for Be at Keck I with the new, improved HIRES.  The original HIRES
(Vogt et al.~1994) has been modified to contain 3 CCDs (blue, green, red)
which combined have 4096 x 6144 pixels and which are 15 $\mu$ in size.  Most
importantly for this work the 8\% quantum efficiency (QE) at 3130 \AA\ of the
old TEK chip is now 94\% in the new blue CCD.

\section{Characteristics of the Ultra-Li-Deficient Stars}

Figure 1 shows the halo Li plateau for stars with [Fe/H] $\lesssim$$-$1.0 with
Li abundances (A(Li) = log N(Li) - log N(H) + 12.00) from Boesgaard et
al.~(2005), Novicki (2005), Ryan et al.~(1996, 1999, 2001a) and Guti\'errez et
al.~(1999).  These Li abundances have all been put on the same temperature
scale as discussed by Novicki (2005).  The nine ultra-Li-deficient stars are
shown with their upper limits on A(Li) as inverted solid triangles.  (Other Li
upper limits for three metal-poor stars from Thorburn 1994 are inverted open
triangles.)  The six stars with Li upper limits observed for Be in this work
are the circled filled triangles.  An additional one of the nine stars, G
186-26, has been observed at Subaru (Boesgaard \& Novicki 2005), and is
indicated by a square around the solid triangle.  The stars plotted all have
T$_{\rm eff}$ $>$5700 K to eliminate stars with possible Li depletion from
deeper convection in the lower mass stars.  (The two ultra-Li-deficient stars
which are not observed here are CD $-$31 19466, which is at too large an air
mass to be observed for Be from Mauna Kea, and G 122-69, which is fainter than
the other stars and would require a long integration time for the high
signal-to-noise spectra needed for meaningful detections or upper limits on
Be.)

Most halo, metal-deficient dwarf and turn-off stars have A(Li) $\sim$2.1, but
there is a subset that are Li-deficient as shown in Figure 1. 
The Li-deficiencies were discovered by several different authors.  Spite et
al.~(1984) found an upper limit in A(Li) for HD 97916 of $<$1.2.  Hobbs et
al.~(1991) found that G 186-26 is severely Li depleted with A(Li) $<$1.23.
Thorburn (1992) first reported on the Li-deficiencies in G 122-69 and G 139-8;
the upper limits on A(Li) given in Thorburn (1994) are $<$1.10 for G 122-69,
$<$1.39 for G 139-8 and $<$1.08 for G 186-26.  (She also identified three
other Li-deficient stars with T$_{\rm eff}$ above 5800 K, but they have V
magnitudes of 12.84, 14.4, and 14.24, making them too faint for Be
observations which require high-resolution; the upper limits on Li for these
stars are not very stringent, only a factor of two or so below the halo star
Li plateau (Thorburn 1994 and Thorburn \& Beers 1993)).  Hobbs \& Mathieu
(1991) found an upper limit of A(Li) $\leq$1.75 for BD +25 1981 and of
$\leq$1.69 for G 202-65, while identifying them as blue stragglers.  The halo
dwarf, G 66-30, was found to be deficient in Li by Spite et al.~(1993) with
A(Li) $\leq$1.45; the Ryan et al.~(2001a) limit for this star is $<$1.61.  Two
other Li-deficient halo stars were discovered by Ryan et al.~(2001a); these
are BD +51 1817 with A(Li) $<$1.64 and CD $-$31 19466 with A(Li) $<$1.49.

Various characteristics of these stars have been studied in order to determine
the cause of the Li deficiencies.  The nine halo stars that are
super-deficient in Li are not a homogeneous group.  Several properties are
summarized in Ryan et al.~(2001a).  Four are long-period, single-lined
spectroscopic binaries with periods ranging from 167-688 days (Carney et
al.~(2001, 2005).  Those stars are HD 97916 (P=664d), BD +51 1817 (516d), G
66-30 (688d), and G 202-65 (167d).  Four are thought to be single stars: BD
+25 1981, G 122-69, G 139-8, and G 186-26, while the binarity of CD $-$31
19466 has not been determined.  Elliott \& Ryan (2005) have measured $v$ sin
$i$ in six of the stars, finding only two with very sharp lines (G 139-8 and G
186-26).   Four others, HD 97916, G 202-65, G 66-30, and G 122-69, show
enhanced rotation with values from 4.7 to 10.4 km s$^{-1}$.  Carney et
al.~(2005) also give values for $v$ sin $i$ for four stars (but for the three
that overlap with Elliott \& Ryan (2005) and Ryan et al.~(2002) their values
are 3-4 km s$^{-1}$ higher); for an additional star, BD +51 1817, they find
enhanced rotation of 9 km s$^{-1}$.

For several stars the abundances of elements other than Fe have been
determined.  Norris et al.~(1997) determined abundances of 14 elements in G
66-30, G 139-8 and G 186-26; Ryan et al.~(1998) added G 122-69 to that list.
And Elliott \& Ryan (2005) added HD 97916 and G 202-65.  There is no
consistent set of abnormal abundances.  Soubiran \& Girard (2005) found
abundances of 9 elements in HD 97916; this star has also had its composition
studied by Fulbright (2000) and Chen et al.~(2000).  Elliott \& Ryan (2005)
presented abundances of 8 elements in HD 97916, G 202-65, G 66-30, G 186-26, G
139-8, and G 122-69.  BD +25 1981 was investigated by Peterson (1981) and
Gehren et al.~(2006).  Although the ultra-Li-deficient stars have a range in
[Fe/H] from $-$1.0 to $-$2.8, they have no pattern of anomalous abundances
relative to Fe.  According to Sneden et al.~(2003), blue stragglers ought to
be enhanced in the s-process elements, Sr, Ba, and Pb.  Indeed G 186-26 has
high [Ba/Fe] = +1.0 and [Sr/Fe] = +0.5, but G 139-8 is deficient in both Ba
and Sr (Elliott \& Ryan 2005).

\section{Keck/HIRES Observations}

The spectra for this research were obtained with the new, improved version of
HIRES on the Keck I telescope on four nights in 2004 and 2005.  The
spectrometer now has a three-chip mosaic of CCDs from MIT-Lincoln Labs with 15
$\mu$ pixels.  There is a dramatic improvement in the quantum efficiency on
the blue/ultraviolet chip.  The Q.E.~in the region of the Be II lines at 3130
\AA\ is 94 \%, a 12-fold enhancement over the previous TEK chip.

The binning used on the CCD was X2-Y1; this binning in the direction of the
dispersion was done to optimize the sampling and the readout time.  The
spectral resolution is 0.0134 \AA\ pix$^{-1}$.  We used the UV cross-disperser
and the B5 decker, which is 3.5 x 0.861 arcsec in dimension on the sky.  All
three CCDs, blue, green, and red, were exposed and flat fields of different
exposure times were obtained to match each of the three chips: 50 sec for the
blue/UV, 3-5 sec for the green and 1 sec for the red.  Seven to 11 bias
frames of 0 sec were taken and Th-Ar comparison spectra were obtained at the
beginning and end of each night.

The log of the observations, including the exposure times and signal-to-noise
ratios per pixel (S/N), are given in Table 1.  Cirrus clouds, sometimes thick,
were present on the nights of 07 November 2004 and 15 May 2005.  In order to
obtain meaningful measurements or upper limits on the Be abundance, high S/N
spectra were required for this program because the Be lines could be expected
to be very weak.  Also in Table 1 are three comparison stars that have normal
Li abundances but are similar in [Fe/H] to the six Li-deficient stars.

The data reduction was done in IRAF following standard procedures.  Master
flat fields were made (median-combined) from the seven quartz-lamp exposures.
A master bias frame was similarly made from the 7-11 bias frames.  These were
applied to the stellar frames and the spectral orders were extracted to
produce the 1-D spectra.  Dispersion solutions were made from the Th-Ar
comparison spectra with low-order Legendre polynomial fits to hundreds of
lines.  To minimize the cosmic ray hits none of the individual exposures
exceeded 30 minutes; most of the stars had multiple exposures and these were
co-added using the wavelength shifts found in the cross-correlation techniques
in IRAF.  For one star (BD +25 1981) spectra were obtained on two different
nights and they were co-added via the same technique.

Figures 2-4 show the Be region in a Li-deficient star and a Li-normal star in
pairs that are approximately matched in temperature and [Fe/H].  Figure 2
compares the Be II region in the Li-deficient star, BD +51 1817, with the
Li-normal star, HD 194598.  These stars have similar values of [Fe/H], $-$1.10
and $-$1.23, but there are two noticeable differences: the absence of the Be
II lines and the higher line broadening in the Li-deficient star.  The same
Li-normal star (HD 194598 at [Fe/H] = $-$1.23) was used as a comparison for
the Li-deficient stars, BD +25 1981 with [Fe/H] = $-$1.26 and HD 97916 with
[Fe/H] = $-$0.96.  While BD +25 1981 seems to have extremely weak Be II lines,
HD 97916 seems to have no Be present.

Figure 3 shows the Li- Be-normal star, HD 233511, at [Fe/H] = $-$1.70 and the
Li-deficient star, G 66-30, at [Fe/H] = $-$1.57.  The two remarkable
differences seen in Figure 2 are also seen in Figure 3: the absence of the Be
II lines and the higher line-broadening in the Li-deficient star.  HD 233511
is also a good, Li-normal, comparison star for G 202-65 which has [Fe/H] =
$-$1.50.  The absence of the Be II lines in G 202-65 is pronounced.  Figure 4
shows the difference in spectra in the Be II region of two other metal-poor
stars, one Li-deficient and one Li-normal.  The Be II lines are readily
apparent in the Li-normal star, HD 24289 at [Fe/H] = $-$2.19, but they are
very weak, if present, in the Li-deficient star, G 139-8, at [Fe/H] =
$-$2.26.

The syntheses of the spectra to derive Be abundances or upper limits are
presented in the next section, but this set of comparisons with the Li-normal
stars demonstrates the reality of the Be-deficiencies in the
ultra-Li-deficient stars.

\section{Beryllium Abundances}

The abundance of Be is not very sensitive to the effective temperature or
[Fe/H], but is does depend rather sensitively on the value of log g.  Our
stars have been observed by several authors, primarily to determine their Li
abundances (the results of which are virtually independent of log g).  The
parameters from several studies have been adopted or averaged.  These
parameters are listed in Table 2 along with the references used for each star.
The error estimates on these parameters are based on the general agreement
among the various references, but are at least 40 K in T$_{\rm eff}$ and at
least 0.2 dex in log g.

Of the six ultra-Li-deficient stars, only one (G 202-65) is in the Li-Be dip
region found in the Hyades and other open clusters (e.g. Boesgaard \& Tripicco
1986, Boesgaard \& King 2000).  One star is hotter and four are cooler.  The
coolest star, G 139-8 at 5984 K, is hotter than the region where Li depletion
due to convective mixing sets in.  The Ryan et al.~(2001b) results for A(Li)
vs.~effective temperature with cool stars cuts off at 5600 K and 6000 K to
avoid including cool stars that may have undergone some Li depletion.
Boesgaard et al.~(2005) show that Li depletion sets in near 5600 K for low
metallicity stars ([Fe/H] $\sim$$-$1.7 and near 5800 K for halo stars with
higher [Fe/H] ($\sim$$-$1.4) values.  None of the stars observed here is that
cool. 

The three comparison stars have been previously studied for Li and found to be
undepleted, normal-Li-plateau, halo stars.  The parameters of Ryan and
Deliyannis (1998) were adopted for HD 24289; they find [Fe/H] = $-$2.22 and
A(Li) = 2.19.  For the other two stars the values from Fulbright (2000) were
used.  He finds [Fe/H] = $-$1.23 and A(Li) = 2.20 for HD 194589 and [Fe/H] =
$-$1.70 and A(Li) = 2.17 for HD 233511.  The atmospheric parameters for the
comparison stars are also given in Table 2.

Beryllium abundances or upper limits have been determined through spectrum
synthesis.  The Kurucz (1993) grid of model atmospheres has been used to
interpolate atmospheres with the parameters in Table 2 for each star.  The
latest version of the program MOOG, as modified in 2002 to include the UV
opacity edges, has been applied (Sneden 1973;
http://verdi.as.utexas.edu/moog.html).  With the exception of Be and O, the
abundances of all elements are reduced by the same amount as Fe.  The O
abundance is adjusted to match the OH features, especially the strong one at
3130.570 \AA.  The total region for the synthesis fits covers 3129.5 to 3132.5
\AA.  A smaller region of the fits is shown for each star in Figures 5 - 10.

The three comparison stars were straight-forward to fit and normal Be
abundances at their [Fe/H] values were derived.  For HD 194598 with [Fe/H] =
$-$1.23 the value found for A(Be) = +0.12.  For HD 233511 at $-$1.70 A(Be) was
found to be $-$0.23.  The most metal-poor comparison star, HD 24289, with
[Fe/H] = $-$2.22, has A(Be) = $-$0.83.  (These data points are shown along
with other Li- and Be-normal stars in the final figures in this paper.)

The comparison of the synthetic and observed spectra in Figure 5 shows that
some Be is present in BD +25 1981 which has [Fe/H] = $-$1.26.  (The spectrum
is very different from the Li-normal comparison star, HD 194598, at [Fe/H] =
$-$1.23, shown in Figure 2.)  The stronger Be line at 3130 \AA\ is well fit by
A(Be) = $-$1.00 whereas the weaker line at 3131 \AA\ gives a better match at
$-$0.70.  Giving twice the weight to the result for the stronger line produces
A(Be) = $-$0.90.

In Figure 6 it is apparent that neither Be line is present (the solid line has
no Be) for HD 97916; the upper limit is taken as $-$1.3.  Similarly, there
seems to be no Be in the spectrum of BD +51 1817 in Figure 7.  For this star
the upper limit on A(Be) is $<$$-$1.50.

For G 66-30, shown in Figure 8, there is no evidence of a Be line at 3130
\AA\, but the data do not rule out Be at the 3131 \AA\ line.  The limit of
A(Be) $<-$1.0 is adopted for this star.  A similar situation prevails for G
202-65 shown in Figure 9: no apparent Be feature at 3130 \AA\, but possible
indication of a small feature (noise?) at 3131 \AA.  Again, a limit on A(Be)
of $<-$1.3 is adopted.

The situation is different for G 139-8 shown in Figure 10.  Both Be lines
appear in this star.  The best fit for the stronger line is A(Be) = $-$1.60
and $-$1.30 for the weaker line.  With twice the weight on the result from the
stronger line at 3130 \AA, the adopted Be abundance becomes $-$1.50.  However,
one should note that the upper limit on lithium, A(Li) of $<$1.39, (Thorburn
1994) is among the lowest of the ultra-Li deficient stars.

The observed values of A(Be), both detections and upper limits, are given in
Table 3 in column 10 as A(Be)$_{meas.}$, as well as in the text above.
Paragraph 5 in this section has the values for the three comparison stars.
Figure 11 puts these Be results and upper limits in the context of other Be
abundances in halo stars with normal Li.  For consistency the stars in this
figure are from analyses done by our group at the Institute for Astronomy:
Stephens et al.~(1997), Boesgaard et al.~(1999), Boesgaard (2000), Boesgaard
\& Novicki (2006).  It also includes the results from Primas et al.~(2000a,
2000b) which are on a virtually identical temperature scale.  Implications of
these findings are discussed in the next section.

\section{Analysis and Results}

\subsection{Rotation}

Predictions have been made by Pinsonneault et al.~(1992) for the amount of
depletion of both Li and Be resulting from mixing caused by rotation.  They
calculated depletions for models with different metallicities (Z = 0.001 and
0.0001), for various initial angular momenta, (no rotation, $J_o$ = 49.2, 49.7
and 50.2) and for an array of masses (M = 0.60 to 0.85 M$_{\odot}$), covering
various evolutionary ages for each mass.

From these predictions the maximum amount of Be depletion that might result
from rotationally-induced mixing has been estimated.  The maximum prediction
was used because our stars seem to have little or no Be left in their
atmospheres.  So the tabulations for $J_o$ = 50.2, the maximum for the initial
angular momentum, were used.  Then the appropriate metallicity models were
used; the value Z = 0.001 was a good choice for five of the stars, while Z =
0.0001 was the correct choice for G 139-8.  The effective temperature was used
to zero-in on the model with the approximate mass and age.  The mass for each
star could be estimated from the latest ``Yale'' isochrones (Yi et al.~2004)
given each star's metallicity, gravity, and temperature.  In order to estimate
the {\it maximum} possible Be depletion among the models with the appropriate
temperature, metallicity, mass, and age, the largest depletion was selected.

Table 3 shows the Li and Be depletions found this way.  The parameters of
T$_{\rm eff}$ and [Fe/H] are shown for each star from Table 2.  Then the
closest tabulated T$_{\rm eff}$ for the models in Tables 4C and 5C (for $J_o$
= 50.2) from Pinsonneault et al.~(1992) are listed along with the model mass
and age that gives the greatest Be depletion.  This Be depletion, dBe$^9$, and
the initial Be abundance (which is a function of [Fe/H]) give the predicted,
present Be abundance, A(Be)$_{pred.}$ = A(Be)$_{init.}$ - d(Be$^9$), which is
to be compared to the measured A(Be).  (The explicit calculations for Li and
Be depletions were presented in Pinsonneault et al.~1992, but newer models
have been made by Pinsonneault et al.~1999, 2002.  The newer work of
Pinsonneault et al.~2002 explains that for rapidly rotating stars there could
be a saturation effect in the angular momentum loss.  They state that the Li
depletion would be less than that predicted by Pinsonneault et al.~1992.  The
Be depletion would be predicted to be smaller also in those newer models.  The
Li and Be depletions in Table 3 would be even less with the new models and so
the rotation hypothesis for the ultra-Li-deficient stars would be even more
difficult to reconcile with the large observed depletions.)

Figure 12 is similar to Figure 11, but all the Be abundances from that figure
(in Li-normal stars) are down-sized to small points.  Then the mean A(Be) at
each of the [Fe/H] values of our Li-deficient stars are shown as ``wagon
wheels;'' this shows the appropriate ``initial Be.  A dotted line connects
that to the position corresponding to the amount by which A(Be) would be
reduced by the maximum rotation, according to the predictions.  None of those
dotted lines reaches down to the observed Be or Be upper limits.

A typical error on a Be abundance determination is $\pm$0.15 (see Figure 11
for examples).  All the observed upper limits on Be are lower than the Be that
is the predicted ``maximum amount of Be depletion'' by considerably more than
the 1$\sigma$ errors.  Only in the case of G 139-8 could the rotation
hypothesis be consistent with the observed Be abundance, although this Be
detection is 1.5$\sigma$ below the prediction.

Table 3 also shows the present Li abundance predictions, where the model
predictions of d(Li$^7$) and the initial A(Li) of 2.25 (for a typical halo
star) gives A(Li)$_{pred.}$.  These are all considerably lower than the
observational upper limits also given in Table 3, i.e. the upper limits do not
provide a meaningful test of the model predictions for Li.

In their discussion of metal-poor blue stragglers Carney et al.~(2005) point
to the role of Li abundances in distinguishing between the metal-poor blue
stragglers and the young metal-poor dwarfs.  In this context they point to the
Li gap found in the Hyades by Boesgaard \& Tripicco (1986) in the temperature
range 6400 - 6800 K over which all the stars show severe Li deficiencies.
Five of the ultra-Li-deficient halo stars are in their paper and their Table 2
lists one (G 202-65) as a Li-gap star with T$_{\rm eff}$ of 6560 K.  (Their
revised T$_{\rm eff}$ is 6340 K in their Table 5 -- cooler than the Hyades Li
gap.)  Beryllium depletion has also been found in the Hyades and other
clusters and field stars in this temperature range (e.g. Boesgaard et
al.~2004a, 2004b).  Could this be the explanation for all or some of the stars
with the ultra-Li deficiencies?  The Li gap for metal-poor stars, which have
lower masses for a given T$_{\rm eff}$, could occur in a broader or a
different range in T$_{\rm eff}$.  This explanation seems implausible,
however, (even for G 202-65) as there are many halo stars with these warmer
temperatures that show normal halo Li abundances (e.g. Thorburn 1994, Ryan et
al.~2001a, Boesgaard et al.~2005, Mel\'endez \& Ram\'\i rez 2004).  In
addition, Ryan et al.~(2001a) state that the {\it Hipparcos} parallaxes
exclude the possibility that these ultra-Li-deficient stars are
``descendents'' of Li gap stars.

\subsection{Blue Stragglers}

Traditionally, we think of ``blue stragglers'' as the hot, blue stars in
globular clusters that are still on the main sequence (e.g. Sandage 1953) even
though the other blue stars in the cluster have evolved to red giants and
beyond, and the turn-off from the main sequence is at a significantly lower
mass.  There are field stars also that have blue-straggler characteristics.
According to Preston \& Sneden (2000) most of the blue metal-poor field stars
are binaries and at least half of those are blue stragglers.  They further
conclude that those blue stragglers are formed by mass transfer through the
evolution of the primary during its red giant phase, rather than by mergers or
by coalescence processes that occur in the densely populated globular
clusters.  This view is supported by the work of Carney et al.~(2001, 2005)
who find evidence of mass transfer.  Ryan et al.~(2001a) and Carney et
al.~(2005) also point out that the blue-straggler binaries in their studies
are depleted in Li.

Other evidence shows that blue stragglers have Li deficiencies.  Observations
were made by Pritchet \& Glaspey (1991) of the blue stragglers in the open
cluster, M 67, to search for Li, but they could determine only upper limits.
Two of the stars studied here, G 202-65 and BD +25 1981, were classified as
blue stragglers by Hobbs and Mathieu (1991) based on photometry.  They found
these stars to be severely depleted in Li ($\leq$1.69 and $\leq$1.75 dex
respectively).  The work in this paper reveals them to be depleted in Be also
($\leq$$-$1.3 and $-$0.90 respectively).

A secondary (lower mass) star can have accreted mass from a companion that is
now a white dwarf, although that secondary has yet to evolve beyond the main
sequence, or even be a turn-off star.  It would have become more massive via
mass transfer from the primary, but not massive enough to be hotter or bluer
than other main sequence stars or turn-off stars.  Ryan et al.~(2001a) call
them ``blue-stragglers-to-be.''

There are a few distinctions in the methods by which the atmosphere of a blue
straggler could become deficient in Li and Be.  (1) A star we observe now was
initially of low enough mass to have destroyed Li and Be during its {\it
pre}-main-sequence evolution.  It has gained mass (also deficient in Li and
Be) now from mass-transfer during the post-main sequence evolution of its
companion.  It is higher up on the main sequence, but not above the turn-off.
(2) During the evolution of the primary, mass was transferred to the
secondary.  The mass transfer process itself may induce extensive mixing which
would result in the depletion of Li and Be.  (3) The post-main sequence
evolution of the primary star included mass transfer of Li and Be diluted (and
depleted?) material onto the surface of the stars we now observe to be
ultra-Li-deficient and Be-deficient.

Higher values for $v$ sin $i$ have been found for the blue-metal-poor stars
that are binaries by Preston \& Sneden (2000) and Carney et al.~(2005)
relative to the constant-velocity stars at similar temperatures.  Four
ultra-Li-deficient stars (HD 97916, BD +51 1817, G 66-30 and G 202-65) have
long periods and large orbital separations now which imply that the high $v$
sin $i$ values are not the result of tidal effects.  More likely, they are
caused by spin-up during the mass transfer which resulted in their becoming
blue stragglers as Carney et al.~(2005) suggest.  The rapid rotation of the
ultra-Li-deficient stars and their connection to blue stragglers was promoted
by Ryan et al.~(2002).  As Figures 2 and 3 show, the lines in the
ultra-Li-deficient stars BD +51 1817 and G 66-30 are noticeably broader than
their Li-normal comparison stars.  Figures 5-10 show that of the six
ultra-Li-deficient stars studied here, only G 139-8 has sharp lines.  This
ties the Li-deficient/Be-deficient stars to the blue straggler phenomenon
also.

A picture is emerging of mass transfer in a blue straggler that results in a
deficiency in both Li and Be.  It is likely that the field blue stragglers are
caused by mass transfer as put forward by Preston \& Sneden (2000) rather than
binary coalescence.  The mass donor, now a white dwarf, would have transferred
mass as it evolved through its red giant phase.  While becoming a red giant,
its convection zone deepens, and whatever Li and Be remained in its surface
layers when it left the main sequence would have been diluted through mixing
in the entire convection zone.  Typically, a surface convection zone of a few
percent of the stellar mass in a main sequence star grows to become some 60
percent of the mass in the red giant (Iben 1965).  At the very least the
deposited material would have diluted whatever Li and Be had not been
destroyed during earlier evolution (Iben 1965).  Carney et al.~(2005) suggest
that the former primaries are now white dwarf stars with a narrow mass range.
Further support for mass transfer from an evolved star comes from the presence
of neutron-capture elements in the ultra-Li-deficient stars.  Although G
186-26 has enhanced [Ba/Fe], G 139-8 is very low in [Ba/Fe], while other
ultra-Li-deficient stars are normal in Ba and Sr (Ryan et al.~1998).

\section{Summary and Conclusion}

There are nine halo stars that are very deficient in Li with upper limits that
are well below the Li plateau.  That plateau shows a near constant Li
abundance attributed to the (somewhat depleted) primordial Li from Big Bang
nucleosynthesis.  We have taken high signal-to-noise, high resolution spectra
to investigate Be in seven of them: one with HDS on Subaru (Boesgaard \&
Novicki 2005) and six with the new HIRES on Keck I.  Three Li-normal stars
with similar metal contents were also observed with HIRES for comparison.

Upper limits or detections of Be have been determined through spectrum
synthesis techniques.  For all seven stars the Be content is well below the
relationship between A(Be) and [Fe/H] found for Li-normal stars, including the
three Li-normal stars that were chosen here at similar [Fe/H] for comparison
with the program stars.

From this study the following conclusions can be drawn:

1) The Li-deficient stars are also Be-deficient and as such they are genuine
   outliers in the A(Li) - [Fe/H] plane and in the A(Li) - T$_{\rm eff}$
   plane.  Therefore they should not be included in any assessment of the
   value of Big-Bang-produced Li.

2. Pinsonneault et al.~(1992) have made calculations to assess the amount of
   Li and Be depletion which results from rotationally-induced mixing.  They
   did this for an array of angular momenta, metallicities, masses, ages,
   temperatures.  With the parameters selected for the stars studied here, it
   was possible to estimate the specific, maximum depletions of Li and Be for
   each star that would be caused by mixing due to rotation.  For these six
   stars the predicted Be depletions ranged between $-$0.56 to $-$0.64 dex.
   However, the observed {\it differences} from the initial Be (at each
   [Fe/H]) to the currently measured (or upper limit) Be range from $-$0.70 to
   $-$1.90 dex.  Thus the depletions that were measured exceeded those that
   were predicted.  A star-by-star evaluation shows that the predictions fail
   by 2$\sigma$ to 9$\sigma$ to match the observed Be abundances and upper
   limits.

3. The star that comes closest in its Be abundance to the predictions of
   rotationally-induced mixing for the Li and Be depletion is G 139-8.  This
   star has apparently retained some Be.  The measured Be (not an upper limit)
   is $-$1.50 $\pm$0.16 which is 0.70 dex below the mean A(Be) of $-$0.80 at
   [Fe/H] = $-$2.26.  The predicted maximum depletion due to rotational mixing
   would make its A(Be) = $-$1.24, compared to the measured $-$1.50; this star
   is only 1.5$\sigma$ below the {\it maximum} predicted for it.  Its A(Be)
   falls below the lower envelope of the spread in Be at its Fe (see Figure
   11).  Could this mean that some of the observed spread in Be is the
   result of a range in initial angular momenta?  Some of the observed spread
   in A(Be) may be due to non-uniform formation of Be throughout the halo and
   disk, e.g. in the neighborhood of hypernovae (see Boesgaard \& Novicki
   2006), and perhaps some could be due to Be depletion from
   rotationally-induced mixing.

4. The blue-straggler analog is best fit to explain the large Li and Be
   deficiencies found in the seven ultra-Li-deficient stars.  The method of
   creating a blue straggler via mass transfer with the Preston \& Sneden
   (2000) model, would readily cause the deficiencies in those two elements.
   During the evolution of the primary whatever Li and Be remained in the
   atmosphere of the secondary would become diluted by the expansion of the
   surface convection zone and the mixing therein as advanced first by Iben
   (1965).  Furthermore, both G 202-65 and BD +25 1981 are known blue
   stragglers.  At least four are long-period, single-lined spectroscopic
   binaries (HD 97916, BD +51 1817, G 66-30 and G 202-65).  All four of those
   and BD +25 1981 have higher $v$ sin $i$ than normal which is attributed to
   spin-up during mass transfer by Carney et al.~(2005).  The donor of the
   mass is now a white dwarf.  The ultra-Li-Be-deficient stars, that are
   apparently single stars, could be the results of stellar mergers.

Of the nine known ultra-Li-deficient stars seven have now been found to be
Be-deficient also.  In other ways these stars are a disparate group as first
pointed out by Norris et al.~(1997) and Ryan et al.~(1998) based on the
composition of four of them, and augmented with two more stars by Elliott \&
Ryan (2005).  They are a homogeneous group in their deficiencies in both Li
and Be.  The evidence accumulated thus far supports their connection to the
blue straggler phenomenon.

\acknowledgements I am grateful to the Keck Observatory support astronomers,
Grant Hill, Jeffrey Mader, and Hien Tran, for their assistance in the HIRES
set-up and to C.J.~Ma, Megan Novicki, and Jeffrey Rich in the data reduction.
This work was supported by NSF AST 00-97945 and AST 05-05899 to AMB.

\begin{figure}
\plotone{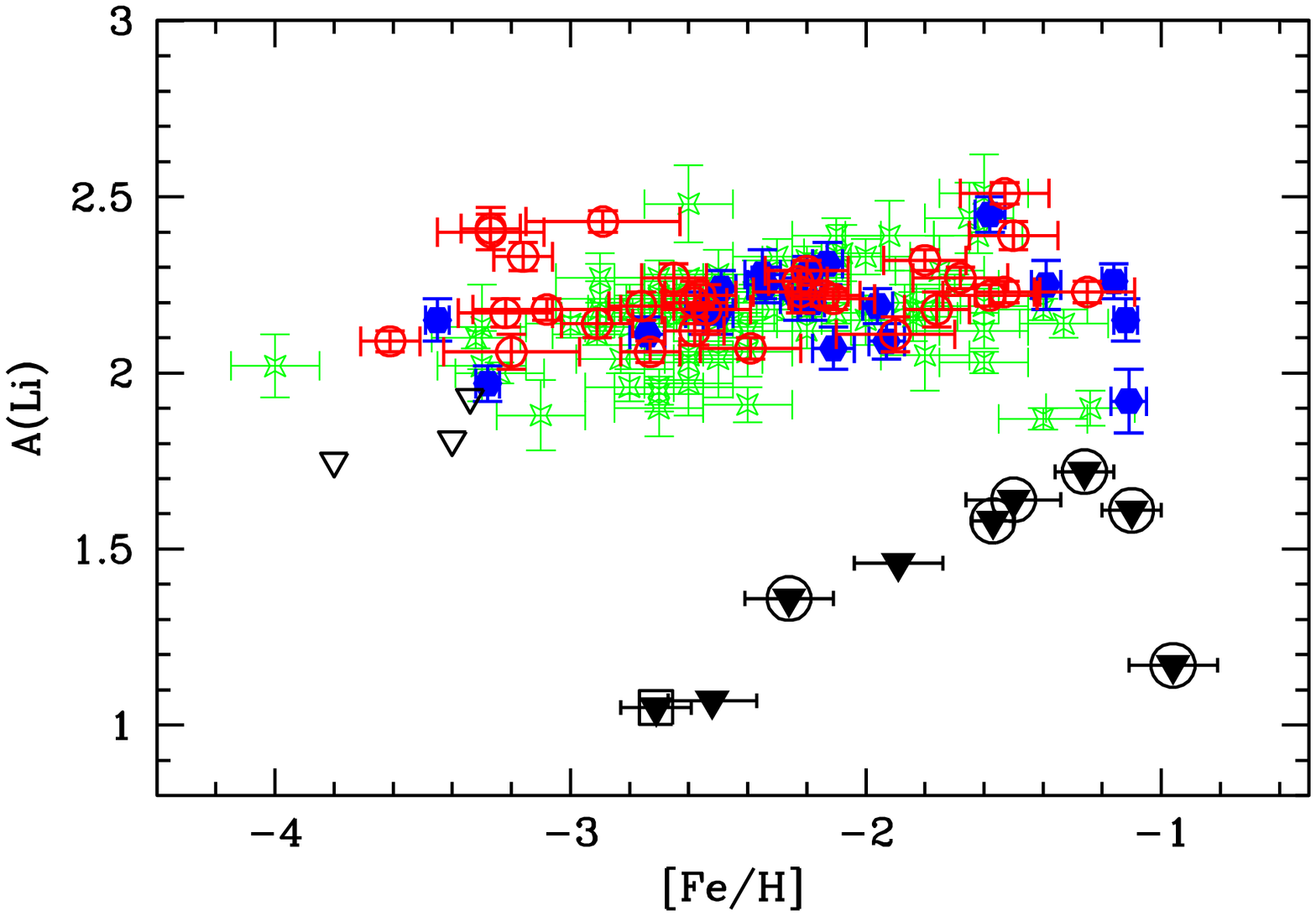}
\caption{The abundance of Li, A(Li) = log N(Li) - log N(H) +12.00 plotted
against [Fe/H] for halo dwarfs.  The open circles are Li abundances from
Novicki (2005), the solid hexagons are from Boesgaard et al.~(2005) for
T$_{\rm eff}$ $>$ 5800 K, the starred squares are from the literature adjusted
to the temperature scale used in the preceding two papers and are from Ryan et
al.~(1996, 1999, 2001a), and Guti\'errez et al.~(1999).  The triangles are Li
upper limits: open triangles from Thorburn (1994) and filled triangles from
Spite et al.~(1984, 1993), Thorburn (1992, 1994) Hobbs \& Mathieu (1991), Ryan
et al.~(2001a) (see text for specifics).  The six filled triangles that are
circled are the stars observed here for Be, while the filled triangle with a
square around it is G 186-26 that was studied for Be by Boesgaard \& Novicki
(2005).  }
\end{figure}

\begin{figure}
\plotone{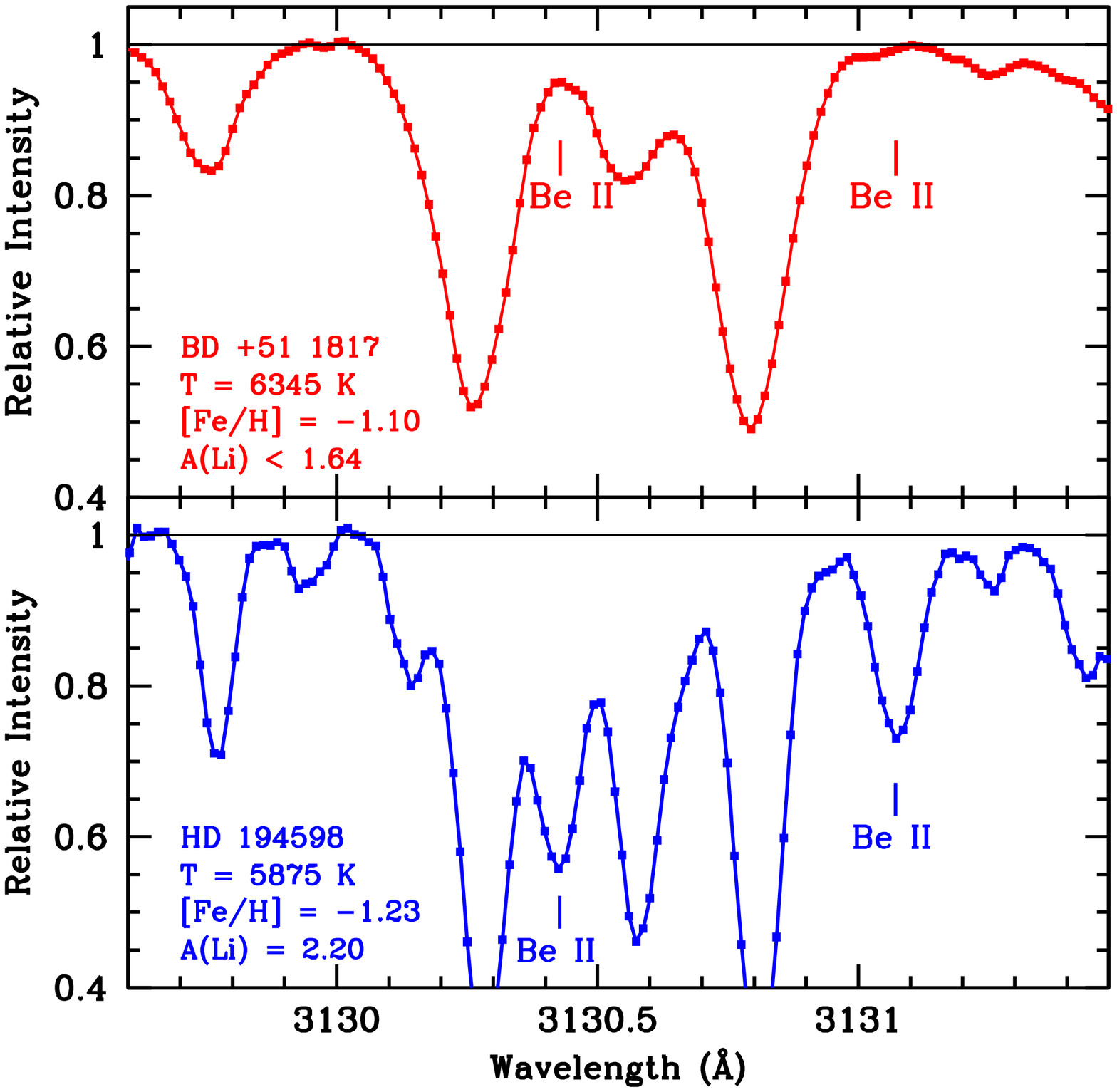}
\caption{The Be II region of the spectra of a Li-deficient star and a
Li-normal star at [Fe/H] $\sim$ $-$1.1.  Note the strong Be II lines in HD
194598 which has A(Li) = 2.20 (Fulbright 2000) and the weak or absent Be II
lines in BD +51 1817 with A(Li) $<$ 1.64 (Ryan et al.~2001a).  BD +51 1817
shows rotational broadening with $v$ $sin$ $i$ = 7.6 $\pm$0.3 from Ryan et
al.~(2002).  (Note that the bottom of the y-axis is 0.40, not 0.00.)}
\end{figure}

\begin{figure}
\plotone{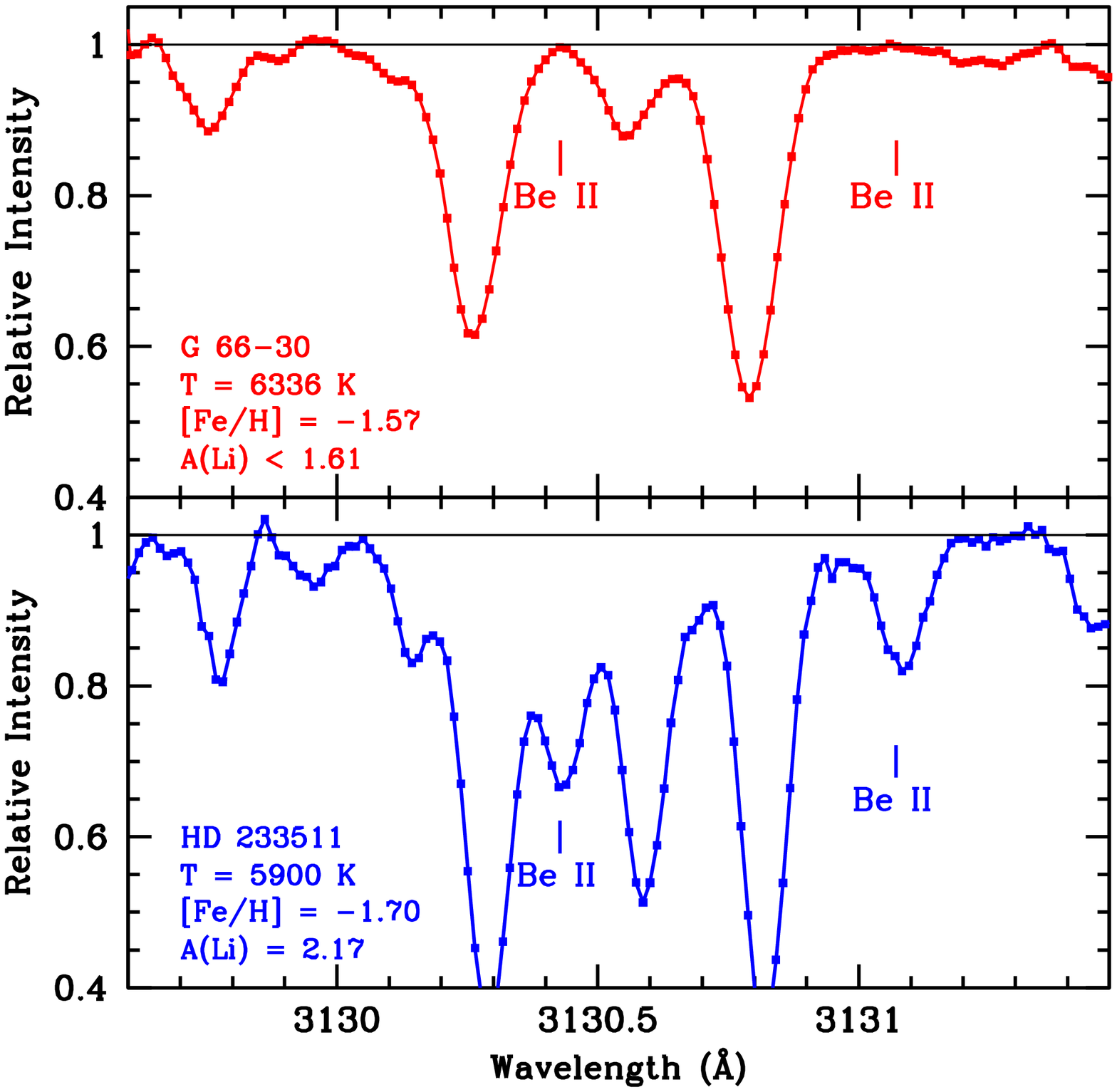}
\caption{The Be II region of the spectra of a Li-deficient star and a
Li-normal star at [Fe/H] $\sim$ $-$1.6.  The Be II lines are strong in the
Li-normal star, HD 233511, which has A(Li) = 2.17 (Fulbright 2000), but appear
to be absent in G 66-30 for which A(Li) $<$ 1.61 (Ryan et al.~2001a).  G 66-30
has rotational broadening with $v$ $sin$ $i$ = 5.5 $\pm$0.6 from Ryan et
al.~(2002).}
\end{figure}

\begin{figure}
\plotone{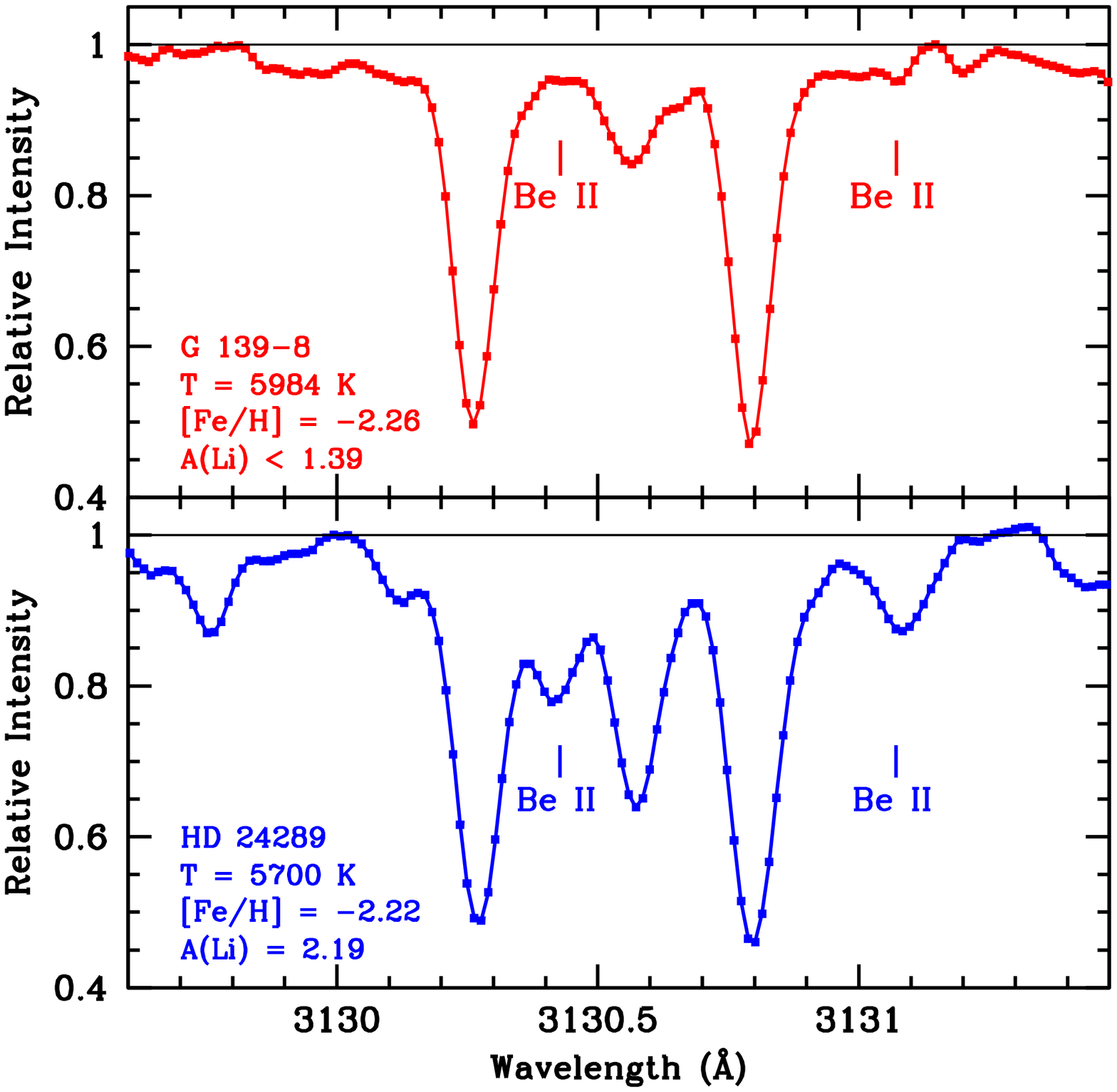}
\caption{The Be II region of the spectra of a Li-deficient star and a
Li-normal star at [Fe/H] $\sim$ $-$2.2.  The Be II lines are clearly present
in HD 24289 which has A(Li) = 2.19 (Ryan \& Deliyannis 1998), while they are
considerably weaker in G 139-8 which has A(Li) $<$ 1.39 (Thorburn 1994).  Both
stars are sharp-lined; the $v$ $sin$ $i$ limit for G 139-8 is $<$1.9 (Elliott
\& Ryan 2005).}
\end{figure}

\begin{figure}
\plotone{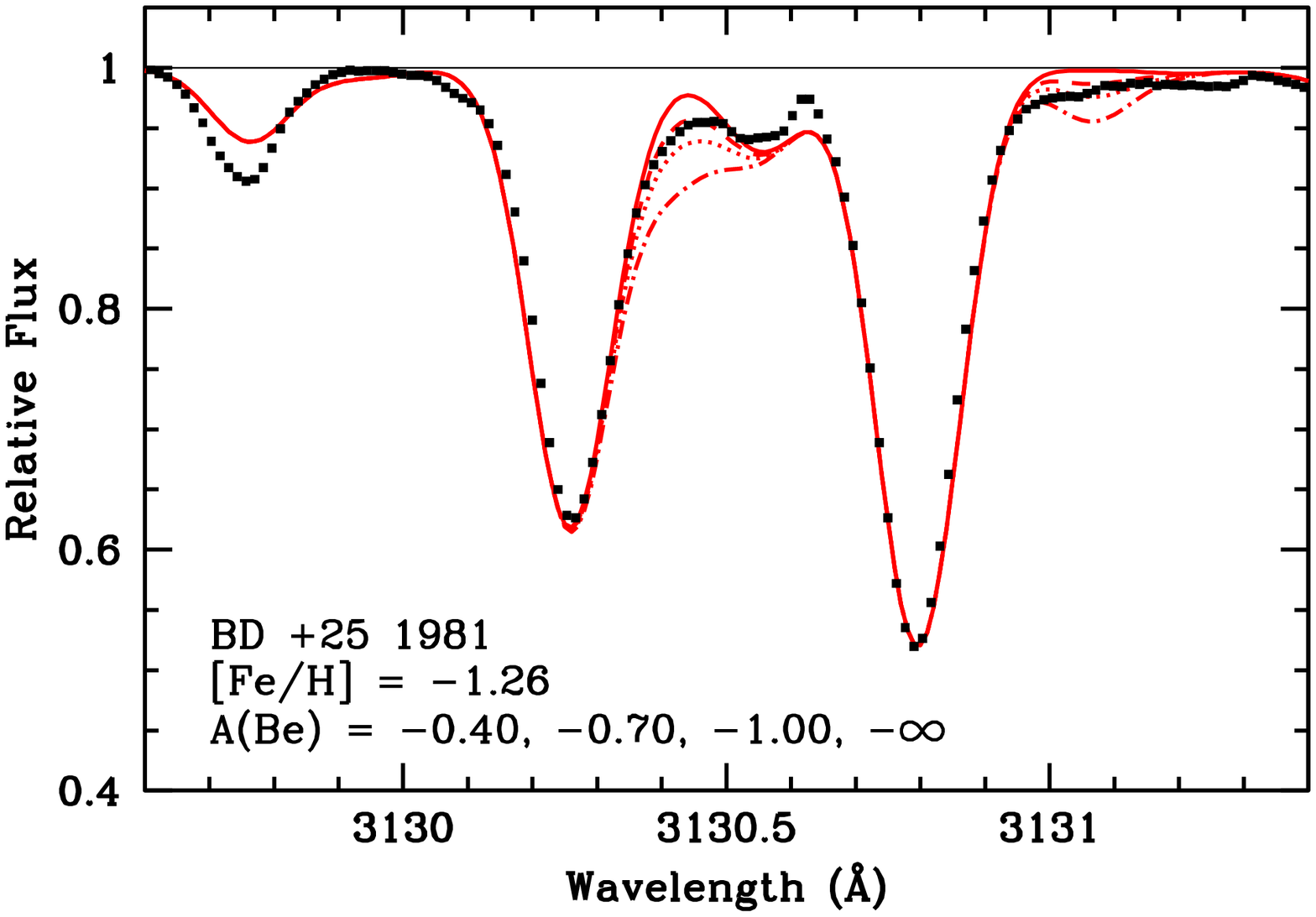}
\caption{Spectrum synthesis fit for BD +25 1981.  The small squares are the
observed points.  The solid line corresponds to no Be, the dashed line to
A(Be) = $-$1.00, the dotted line to $-$0.70, the dot-dash line to $-$0.40.
For the stronger line the best fit appears to be A(Be) = $-$1.0 and for the
weaker line it is $-$0.70.  The stronger line has twice the weight of the
weaker one giving A(Be) = $-$0.90 for BD +25 1981.}
\end{figure}

\begin{figure}
\plotone{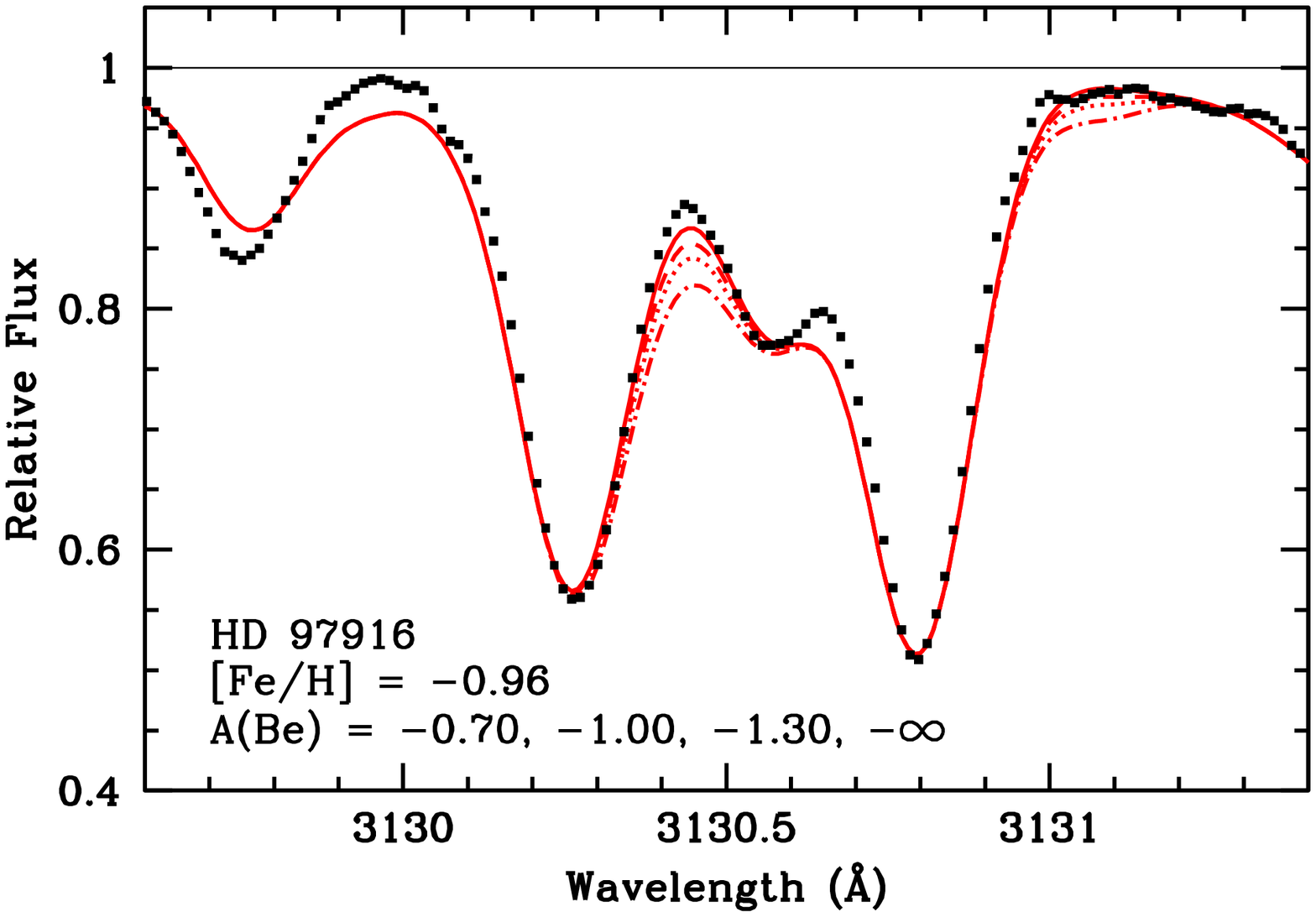}
\caption{Spectrum synthesis fit for HD 97916.  The small squares are the
observed points.  The solid line corresponds to no Be, the dashed line to
A(Be) = $-$1.30, the dotted line to $-$1.00, the dot-dash line to $-$0.70.
The fit is consistent with no Be present; A(Be) $<$ $-$1.0 is adopted as the
upper limit.}
\end{figure}

\begin{figure}
\plotone{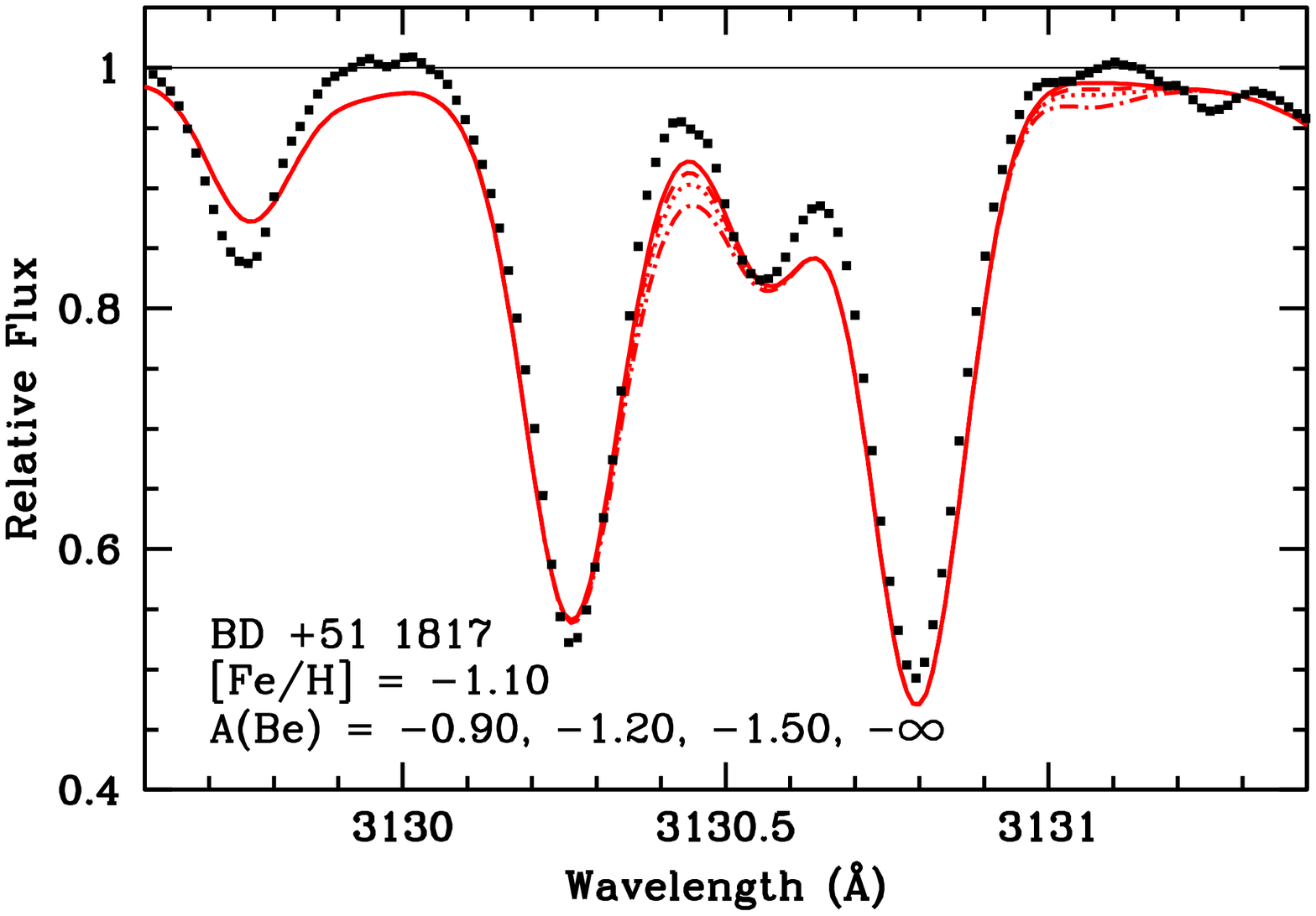}
\caption{Spectrum synthesis fit for BD +51 1817.  The small squares are the
observed points.  The solid line corresponds to no Be, the dashed line to
A(Be) = $-$1.50, the dotted line to $-$1.20, the dot-dash line to $-$0.90.
This star has no evidence of Be and a limit of $-$1.5 is adopted.}
\end{figure}

\begin{figure}
\plotone{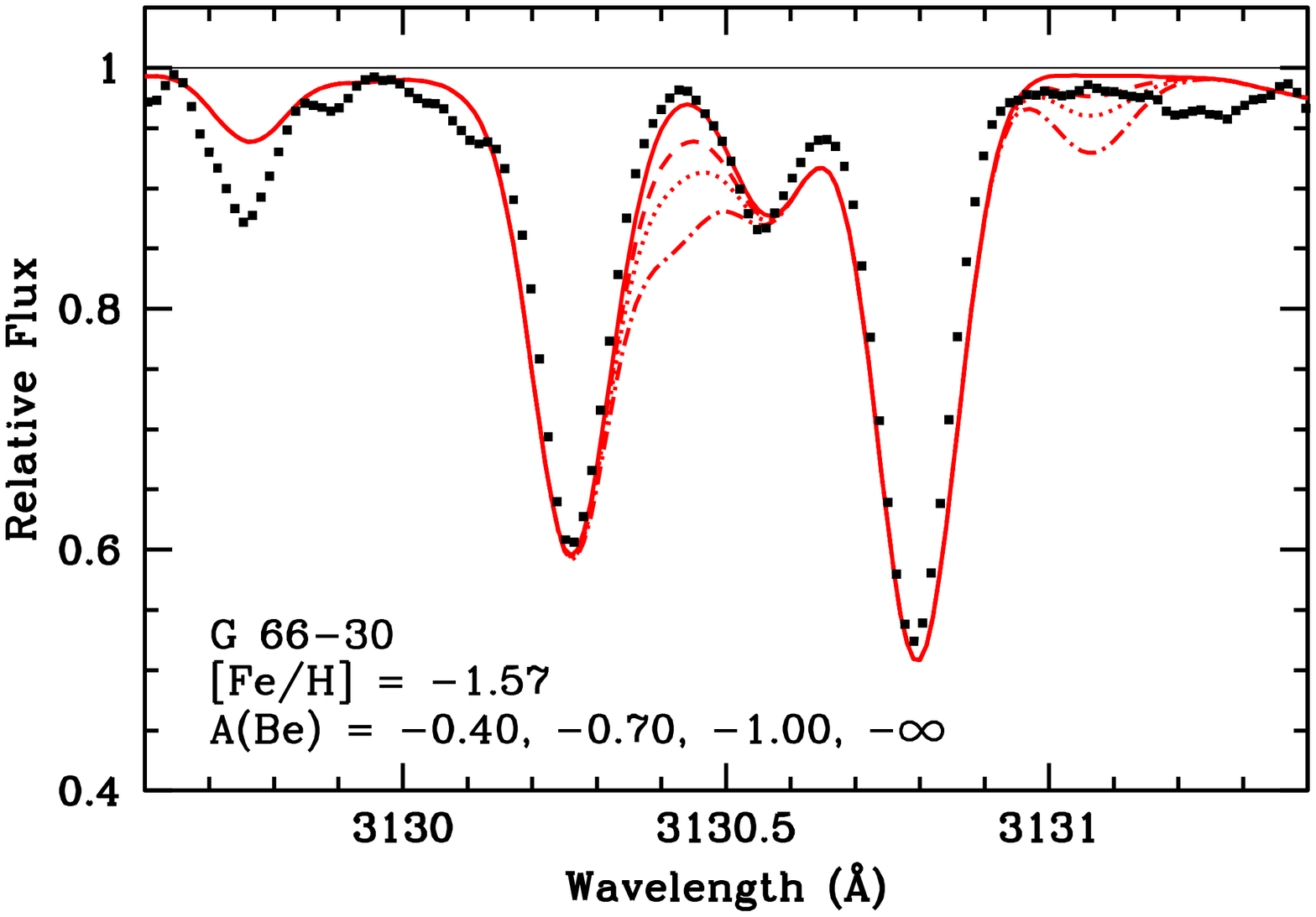}
\caption{Spectrum synthesis fit for G 66-30.  The small squares are the
observed points.  The solid line corresponds to no Be, the dashed line to
A(Be) = $-$1.00, the dotted line to $-$0.70, the dot-dash line to $-$0.40.
The stronger line is consistent with no Be, but the weaker line indicates that
Be may be present at the $-$1.0 level.  An upper limit of A(Be) $<$ $-$1.0 is
adopted.}
\end{figure}

\begin{figure}
\plotone{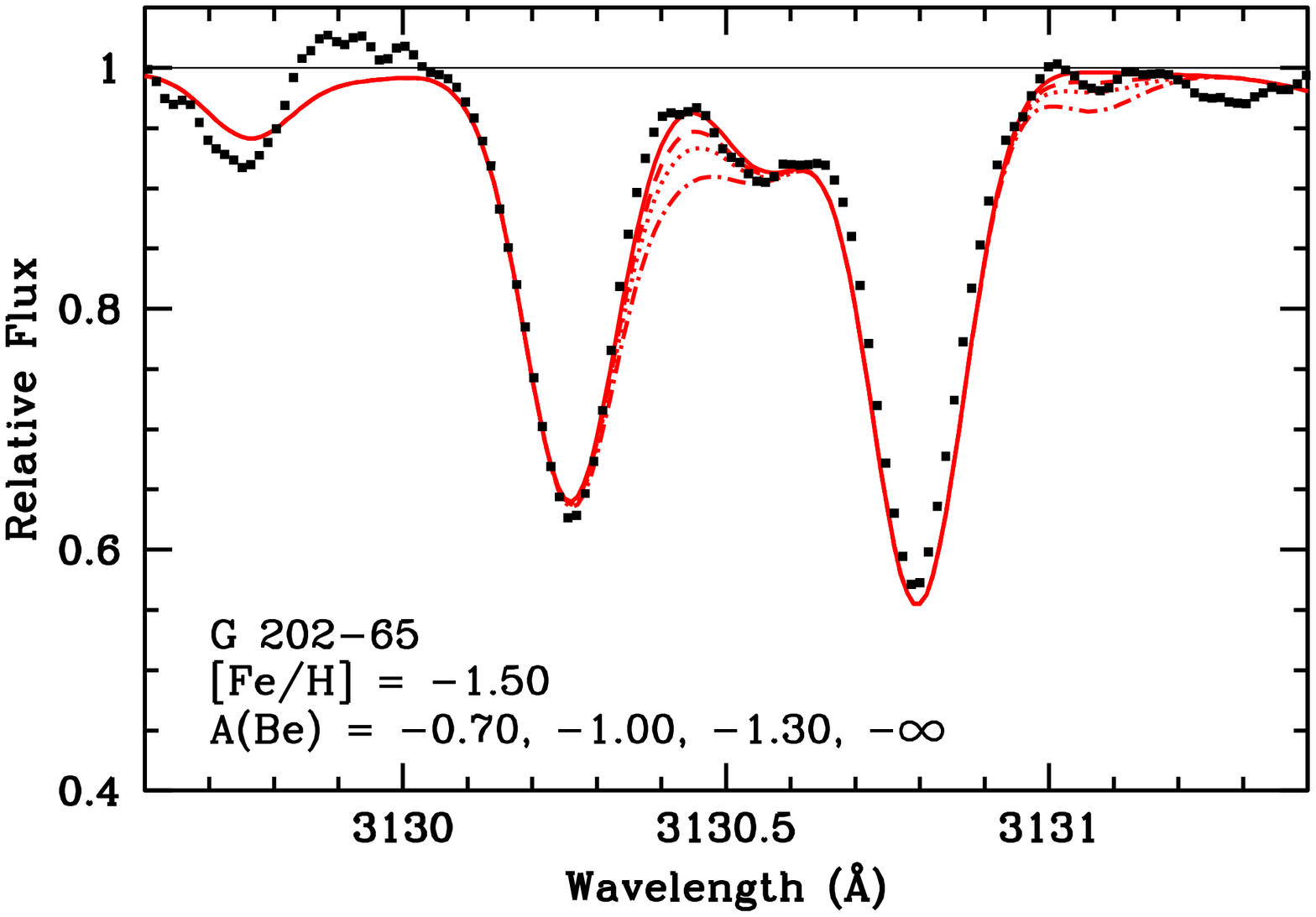}
\caption{Spectrum synthesis fit for G 202-65.  The small squares are the
observed points.  The solid line corresponds to no Be, the dashed line to
A(Be) = $-$1.30, the dotted line to $-$1.00, the dot-dash line to $-$0.70.  An
upper limit of A(Be) $<$ $-$1.3 is adopted for this star, primarily based on
the stronger line.}
\end{figure}

\begin{figure}
\plotone{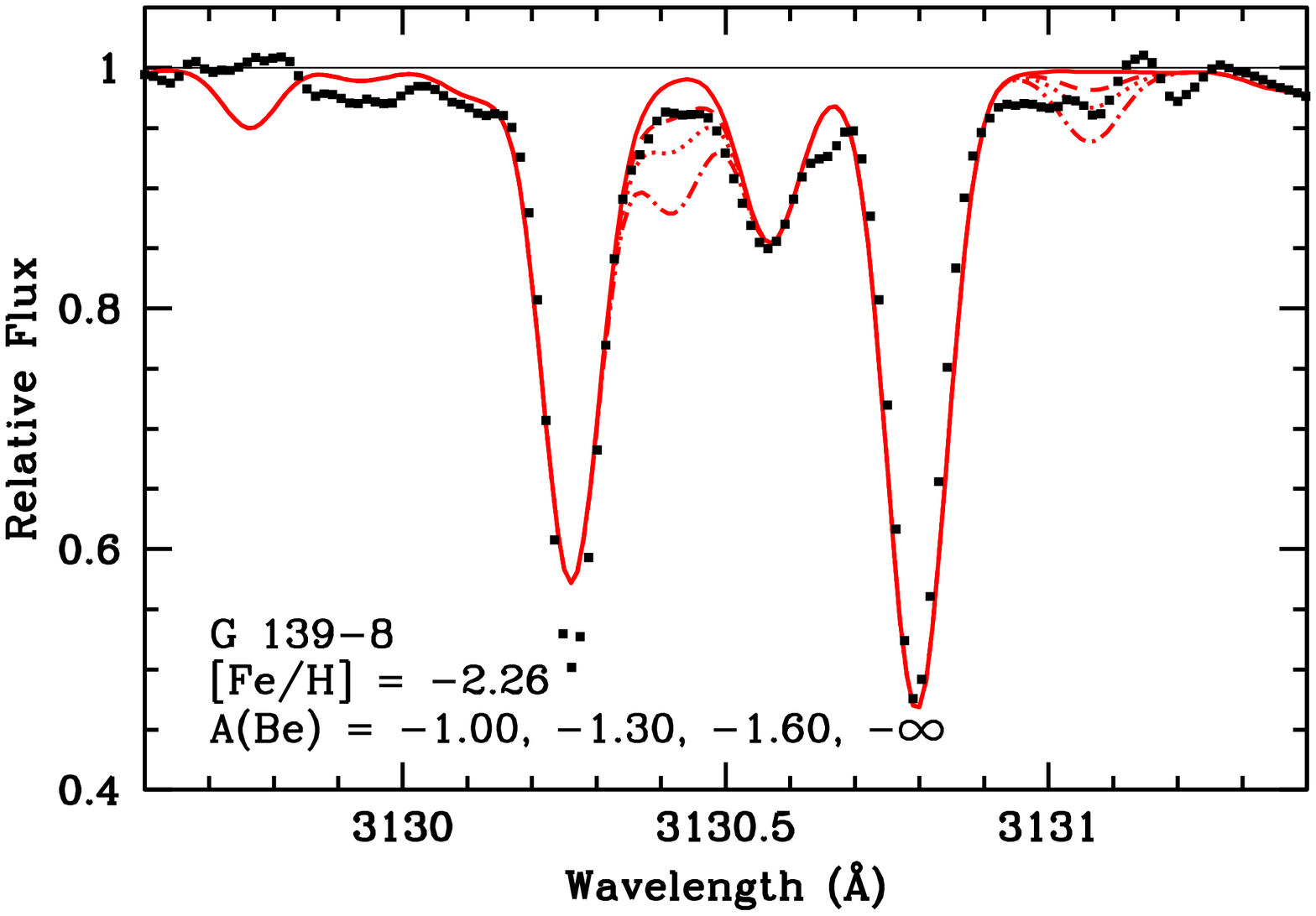}
\caption{Spectrum synthesis fit for G 139-8.  The small squares are the
observed points.  The solid line corresponds to no Be, the dashed line to
A(Be) = $-$1.60, the dotted line to $-$1.30, the dot-dash line to $-$1.00.
This star does appear to have some Be and a value for A(Be) is $-$1.5, based
on $-$1.60 for the stronger line and $-$1.30 for the weaker line.}
\end{figure}

\begin{figure}
\plotone{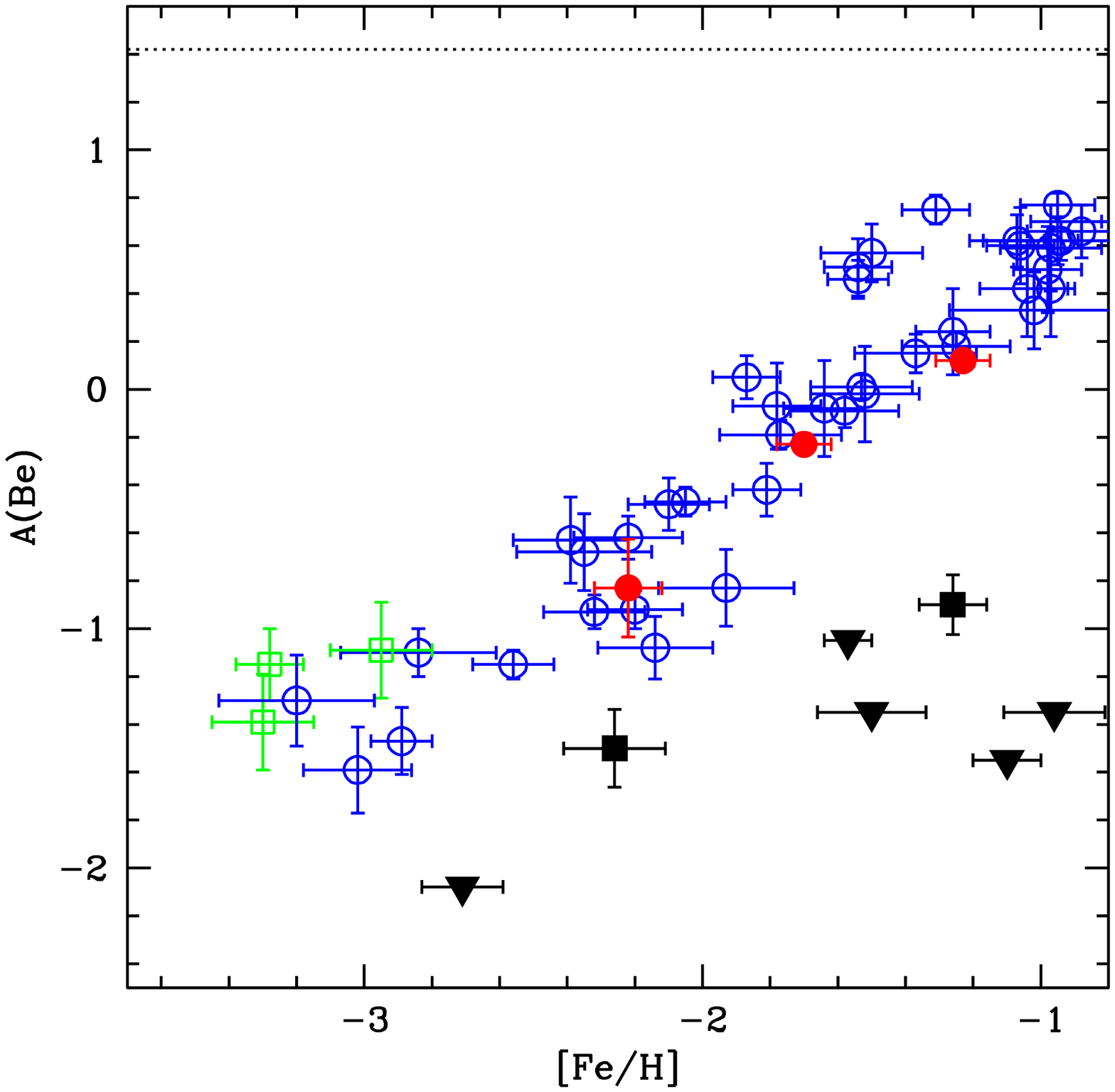}
\caption{The Be abundances/limits of the ultra-Li-deficient stars shown in the
context of Be abundances in Li-normal stars as a function of [Fe/H].  The
filled squares are the two ultra-Li-deficient stars with detected Be while the
filled triangles represent the upper limits on four ultra-Li-deficient stars.
Open circles are from Boesgaard et al.~(1999), Boesgaard (2000), Boesgaard et
al.~(2004).  Filled circles are the Li-normal stars in this paper.  Open
squares (low metallicity stars) are from Primas et al.~(2000a, 2000b).
Individual error bars on both A(Be) and [Fe/H] are shown.  The horizontal
dotted line is the meteoritic Be abundance, A(Be) = 1.42 (Grevesse \& Sauval
1999).}
\end{figure}

\begin{figure}
\plotone{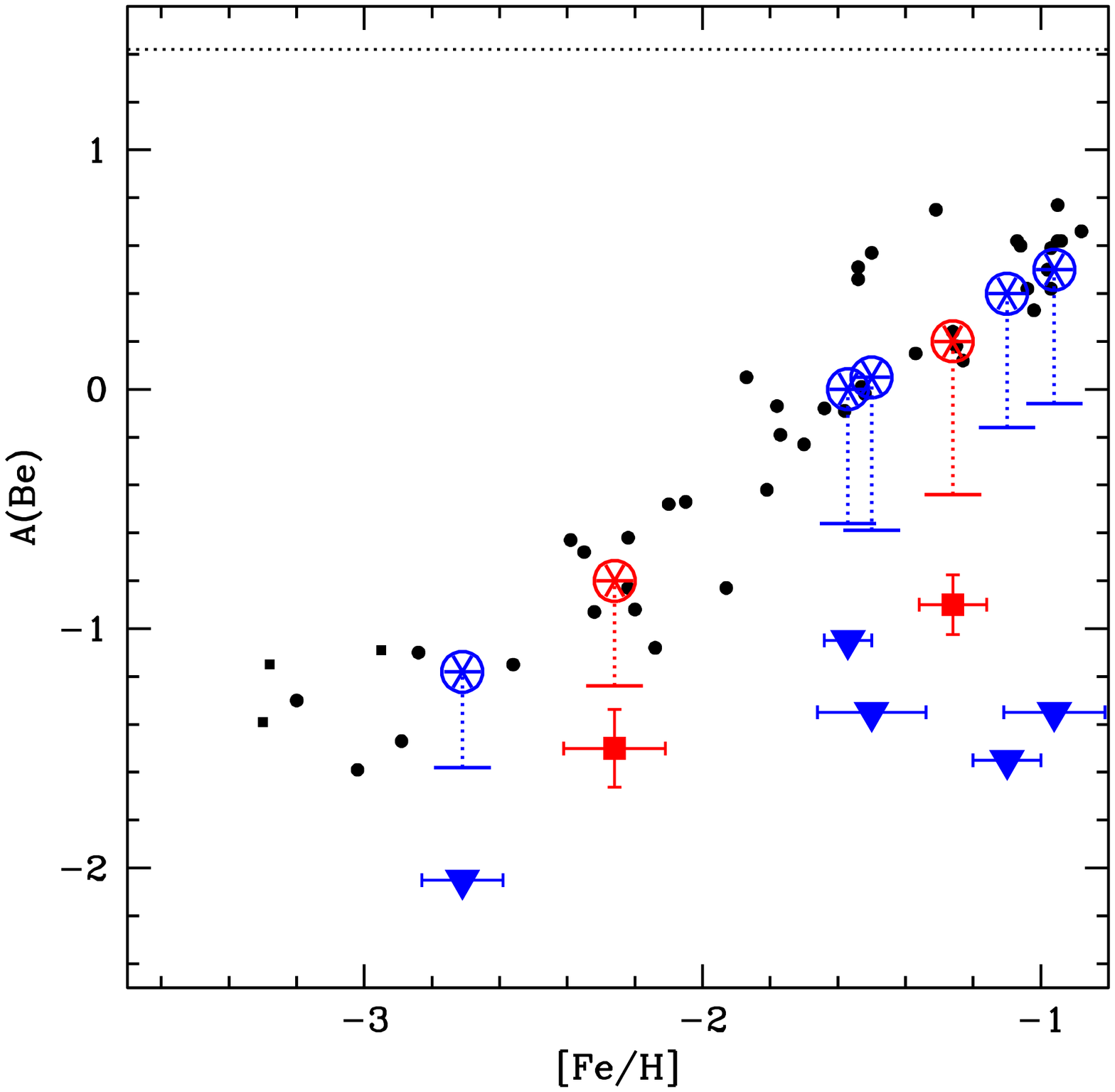}
\caption{Be abundances predicted from the rotation models of Pinsonneault et
al.~(1992).  The symbols representing the Be abundances in the Li-normal stars
in the previous figure have been down-sized as solid symbols.  The
``wagon-wheels'' represent the initial Be abundance for each of the
ultra-Li-deficient stars; the dotted lines emanating downward, ending in a
horizontal line, show the Be abundance as reduced by the maximum amount
predicted from rotationally-induced mixing by Pinsonneault et al.~(1992).
This shows that the rotation models do not predict enough Be depletion to
account for the observations. }
\end{figure}

\clearpage

\singlespace
\begin{center}
\begin{deluxetable}{lrcrcc} 
\tablewidth{0pc}
\tablecolumns{7} 
\tablecaption{Keck HIRES Observations} 
\tablehead{ 
\colhead{Star}  &  \colhead{V}  & \colhead{(B-V)$_{0}$}  & 
\colhead{Night} & \colhead{Exp. Time} & \colhead{Total}  \\

\colhead{} & \colhead{} & \colhead{}  & \colhead{} & \colhead{(min)} & \colhead{S/N}} 
\startdata 
\sidehead{Li-deficient stars:}
BD +25 1981 &  9.29   & 0.36 &  07 Nov 2004  & \phn 30 & 170 \\
BD +25 1981 &  9.29   & 0.36 &  31 Jan 2005  & \phn 30 & 160 \\
HD 97916    &  9.17   & 0.38 &  31 Jan 2005  & \phn 20 & 120 \\
BD +51 1817 & 10.23   & 0.38 &  31 Jan 2005  & \phn 85 & 128 \\
G 66-30     & 11.07   & 0.37 &  01 Apr 2005  & \phn 90 & 120 \\
G 202-65    & 11.22   & 0.36 &  01 Apr 2005  & \phn 90 & \phn 90 \\
G 139-8     & 11.46   & 0.42 &  15 May 2005  &     150 & 102 \\
\sidehead{Li-normal stars:}
HD 24289    & 9.96    & 0.52 &  07 Nov 2004  & \phn 50 & \phn 97 \\
HD 233511   & 9.71    & 0.48 &  31 Jan 2005  & \phn 30 & 118 \\
HD 194598   & 8.36    & 0.49 &  07 Nov 2004  & \phn 10 & 140 \\

\enddata 

\end{deluxetable} 
\end{center}


\clearpage

\begin{center}

\begin{deluxetable}{lccccccccccc} 
\tightenlines
\singlespace
\tabletypesize{\tiny}
\thispagestyle{empty}

\tablewidth{0pc} 
\tablenum{2} 
\tablecolumns{12} 
\tablecaption{Stellar Parameters}
\tablehead{ \colhead{Star} & \colhead{T$_{\rm eff}$} & \colhead{ref.\tablenotemark{1}}
& \colhead{$\sigma$(T)} 
& \colhead{$\log{\rm g}$} & \colhead{ref.\tablenotemark{1}}   & \colhead{$\sigma$(log g)} &
\colhead{[Fe/H]} & 
\colhead{ref.\tablenotemark{1}}   & \colhead{$\sigma$([Fe/H])} &
\colhead{$\xi$} & \colhead{ref.}  \\ 

\colhead{} & \colhead{(K)} & \colhead{} & \colhead{(K)} & \colhead{(dex)} &
\colhead{} & \colhead{(dex)} & \colhead{(dex)} & \colhead{} & \colhead{(dex)}  
& \colhead{(km~s$^{-1}$)} & \colhead{} 
}
\startdata 
\sidehead{Li-deficient stars:}
BD +25 1981 & 6811 & P81 & $\pm$40  & 4.00 & P81 & $\pm$0.40 & $-$1.26 & P81 &
$\pm$0.10 & 1.5 & M84 \\
HD 97916    & 6375 & E05 & $\pm$81  & 3.95 & E05 & $\pm$0.20 & $-$0.96 & E05
 & $\pm$0.15 & 1.6 & E05 \\
BD +51 1817 & 6345 & R01 & $\pm$40  & 4.00 & --- & $\pm$0.40 & $-$1.10 & R01 & 
$\pm$0.10 & 1.5 & M84 \\
G 66-30     & 6336 & R01,E05,N97 & $\pm$90 & 4.20 & E05,N97,S93 & $\pm$0.20 &
$-$1.57 & R01,E05,N87 & $\pm$0.07 & 1.3 & E05 \\
G 202-65    & 6480 & R01,C05,E05,CP05 & $\pm$75 & 3.85 & CP05 & $\pm$0.20 & 
$-$1.50 & R01 & $\pm$0.16 & 1.55 & E05 \\
G 139-8     & 5984 & CP05 & $\pm$40  & 3.50 & CP05 & $\pm$0.40 & $-$2.26 &
E05,N97 & $\pm$0.15 & 0.9 & E05 \\
\sidehead{Li-normal stars:}
HD 24289 & 5700 & RD98 & $\pm$100 & 3.5 & RD98 & $\pm$0.50 & $-$2.22 & RD98 &
$\pm$0.10 & 1.5 & M84 \\
HD 194598 & 5875 & F00 & $\pm$40 & 4.2 & F00 & $\pm$0.06 & $-$1.23 & F00 &
$\pm$0.08 & 1.4 & F00 \\
HD 233511 & 5900 & F00 & $\pm$40 & 4.2 & F00 & $\pm$0.06 & $-$1.70 & F00 &
$\pm$0.08 & 1.2 & F00 \\
\tablenotetext{1}{
C05=Carney, Laird \& Latham 2005;
CP05=Charbonnel \& Primas 2005;
E05=Elliott \& Ryan 2005;
F00=Fulbright 2000;
M84=Magain 1984;
N97=Norris et al.~1997;
P81=Peterson 1981;
RD98=Ryan \& Deliyannis 1998;
R01=Ryan et al.~2001;
S93=Spite et al.~1993}
\enddata 
\end{deluxetable} 
\end{center}

\clearpage

\begin{center}

\begin{deluxetable}{lcccccccccccc} 
\tightenlines

\singlespace
\tabletypesize{\tiny}
\thispagestyle{empty}
\tablewidth{0pc} 
\tablenum{3} 
\tablecolumns{13} 
\tablecaption{Li and Be Depletions}
\tablehead{ \colhead{Star} & \colhead{[Fe/H]} & \colhead{T$_{\rm eff}$} 
& \colhead{model T$_{\rm eff}$\tablenotemark{1}}
& \colhead{Model Mass\tablenotemark{1}}
& \colhead{Model Age\tablenotemark{1}} & \colhead{d(Be$^9$)}   
& \colhead{A(Be)$_{init.}$} 
& \colhead{A(Be)$_{pred.}$} & \colhead{A(Be)$_{meas.}$}   
& \colhead{A(Li)$_{pred.}$} & \colhead{A(Li)$_{meas.}$} 
& \colhead{Li ref.\tablenotemark{2}}  \\ 

\colhead{}  & \colhead{} & \colhead{(K)}  & \colhead{(K)}
& \colhead{(M$_{\odot}$)} & \colhead{(Gyr)} & \colhead{(dex)} &
\colhead{(dex)} & \colhead{(dex)} & \colhead{(dex)} & \colhead{dex} 
& \colhead{(dex)} & \colhead{} \\
}
\startdata 
BD +25 1981 & $-$1.26 & 6811 & 6517 & 0.85  & 10.9 & $-$0.64 & 0.2    
& $-$0.44 & $-$0.90   & 0.53 & $\leq$1.75 & HM91 
 \\
HD 97916    & $-$0.96 & 6375 & 6346 & 0.825 & 10.0 & $-$0.56 & 0.5    
& $-$0.06 & $<$$-$1.3 & 0.72  & $<$1.2 & S84
 \\
BD +51 1817 & $-$1.10 & 6345 & 6346 & 0.825 & 10.0 & $-$0.56 & 0.4    
& $-$0.16 & $<$$-$1.5 & 0.72  & $<$1.64 & R01
 \\
G 66-30     & $-$1.57 & 6336 & 6346 & 0.825 & 10.0 & $-$0.56 & 0.0    
& $-$0.56 & $<$$-$1.0 & 0.72  & $<$1.50 & S93
 \\
G 202-65    & $-$1.50 & 6480 & 6490 & 0.85  & 10.0 & $-$0.64 & 0.0    
& $-$0.64 & $<$$-$1.3 & 0.55  & $<$1.67 & R01
 \\
G 139-8     & $-$2.26 & 5984 & 5956 & 0.775 &  3.0 & $-$0.44 & $-$0.8 
& $-$1.24 & $-$1.50   & 1.02  & $<$1.39 & T94
 \\

\tablenotetext{1}{selected to maximize the Be depletion due to rotation
effects, not meant to be exact match.} 

\tablenotetext{2}{HM91 = Hobbs \& Mathieu 1991; R01 = Ryan et al.~2001;
S84 = Spite et al.~1984; S93 = Spite et al.~1993; T94 = Thorburn 1994}

\enddata 
\end{deluxetable} 
\end{center}

\end{document}